\pdfoutput=1

\documentclass[11pt,twoside,a4paper,cmspaper,final,collab]{cms-tdr}
\def\svnVersion{046db09-D}\def\svnDate{2020/04/06}\def\cmsCernNoTag{CERN-EP-2019-265}\def\cmsCernDate{\today}\def\cmsMessage{"Published in Physical Review Letters as \href{http://dx.doi.org/10.1103/PhysRevLett.124.131802}{\doi{10.1103/PhysRevLett.124.131802}.}"}

\begin{document}\cmsNoteHeader{EXO-19-018}

\hyphenation{had-ron-i-za-tion}
\hyphenation{cal-or-i-me-ter}
\hyphenation{de-vices}
\title{Search for a Narrow Resonance Lighter than \texorpdfstring{200\GeV}{200 GeV} Decaying to a Pair of Muons in Proton-Proton Collisions at \texorpdfstring{$\sqrt{s}=13\TeV$}{sqrt(s) = 13 TeV}}

\date{\today}

\abstract{A search is presented for a narrow resonance decaying to a pair of oppositely charged muons using $\sqrt{s}=13\TeV$ proton-proton collision data recorded at the LHC. In the 45--75 and 110--200\GeV resonance mass ranges, the search is based on conventional triggering and event reconstruction techniques. In the 11.5--45\GeV mass range, the search uses data collected with dimuon triggers with low transverse momentum thresholds, recorded at high rate by storing a reduced amount of trigger-level information. The data correspond to integrated luminosities of 137\fbinv and 96.6\fbinv for conventional and high-rate triggering, respectively. No significant resonant peaks are observed in the probed mass ranges. The search sets the most stringent constraints to date on a dark photon in the $\sim$30--75 and 110--200\GeV mass ranges.}

\hypersetup{
pdfauthor={CMS Collaboration},
pdftitle={Search for a narrow resonance lighter than 200 GeV decaying to a pair of muons in proton-proton collisions at 13 TeV},
pdfsubject={CMS},
pdfkeywords={CMS, physics, dark photon, muon scouting}}

\maketitle

A large body of cosmological evidence points to the existence of dark matter~\cite{Bertone:2010zza,Feng:2010gw,Porter:2011nv,Ade:2015xua}. Unraveling its origin remains one of the outstanding problems of particle physics and cosmology. Dark matter is expected to interact very weakly, if at all, with standard model (SM) particles. This raises the possibility of a hidden, dark sector of particles whose interaction with SM particles may be mediated by a hypothetical dark photon ($\PZ_{D}$)~\cite{Essig:2013lka}.

We present a search for a narrow resonance decaying to a pair of oppositely charged muons using $\sqrt{s}=13\TeV$ proton-proton ($\Pp\Pp$) collision data recorded by the CMS experiment at the CERN LHC. The search looks for a narrow resonance in the 11.5--200\GeV mass range, omitting the 75--110\GeV range where \PZ boson production dominates. The results of this search are interpreted in the context of a $\PZ_{D}$ that interacts with SM particles through the kinetic mixing of its $\mathrm{U}(1)_{D}$ gauge field with the $\mathrm{U}(1)_{Y}$ hypercharge field of the SM~\cite{Galison:1983pa,Holdom:1985ag,Dienes:1996zr}. The degree of mixing and the strength of the coupling of $\PZ_{D}$ to SM particles is determined by the kinetic mixing coefficient $\epsilon$. A discussion of the theory of kinetic mixing and its impact on electroweak symmetry breaking and on the electroweak precision variables can be found in Ref.~\cite{Curtin:2014cca}. Constraints have been placed on visible $\PZ_{D}$ decays in direct searches by beam dump~\cite{Konaka:1986cb}, fixed-target~\cite{Abrahamyan:2011gv}, rare meson decay~\cite{MeijerDrees:1992kd}, and collider~\cite{Lees:2014xha,Ilten:2016tkc,Aaij:2017rft,Aaij:2019bvg} experiments.

The central feature of the CMS apparatus is a superconducting solenoid of 6\unit{m} internal diameter, providing a magnetic field of 3.8\unit{T}. Within the solenoid volume are a silicon pixel and strip tracker, a lead tungstate crystal electromagnetic calorimeter, and a brass and scintillator hadron calorimeter, each composed of a barrel and two end cap sections. Forward calorimeters extend the pseudorapidity ($\eta$) coverage provided by the barrel and end cap detectors. Muons are detected in gas-ionization chambers embedded in the steel flux-return yoke outside the solenoid. A detailed description of the CMS detector, together with a definition of the coordinate system used and the relevant kinematic variables, can be found in Ref.~\cite{Chatrchyan:2008zzk}.

Events of interest are selected using a two-tiered trigger system~\cite{Khachatryan:2016bia}. The first level (L1), composed of custom hardware processors, uses information from the calorimeters and muon detectors to select events at a rate of around 100\unit{kHz} within a time interval of 4\mus. The second level, known as the high-level trigger (HLT), consists of a farm of processors running a version of the full event reconstruction software optimized for fast processing. A set of triggers in the HLT system reduces the event rate to around 1\unit{kHz}, allowing data containing all the information necessary for complete event reconstruction, $\sim$1\unit{MB}/event, to be transferred to storage. These triggers are termed ``standard triggers'' in this Letter. The standard dimuon triggers have transverse momentum (\pt) requirements of 12--15\GeV on the highest \pt muon, and 5--7\GeV on the second-highest \pt muon reconstructed at L1. Requirements of $\pt > 17$ and 8\GeV are imposed at the HLT on the highest and second-highest \pt muons, respectively. The standard dimuon triggers collect data at a rate of around 30\unit{Hz} at the peak instantaneous luminosity of $2\times10^{34}\percms$. These data are then fully reconstructed and used in the search for dimuon resonance masses greater than 45\GeV. Dimuon events that do not pass the standard dimuon triggers, but are selected by certain single-muon triggers are also included. These required $\pt > 24\GeV$ in the 2016 and 2018 data taking periods, and $\pt > 27\GeV$ in 2017. The data collected during the years 2016--2018 correspond to an integrated luminosity of 137\fbinv.

For dimuon resonance masses below $\sim$40\GeV, the \pt thresholds in the standard dimuon triggers significantly reduce the signal acceptance, thereby adversely affecting the search sensitivity. Therefore, a dedicated set of triggers with significantly lower muon \pt thresholds were implemented. Events selected with these triggers contain a very limited amount of information about muons reconstructed at the HLT. This includes the muon four momenta, the number of hits left by each muon track in the tracking system, the normalized $\chi^{2}$ of the muon track, and the isolation around the muon computed as the scalar sum of the \pt of all additional tracks in a $\Delta R = \sqrt{\smash[b]{(\Delta\eta)^2+(\Delta\phi)^2}} = 0.3$ cone around the muon. The resulting event size is $\sim$4\,(8)\unit{kB} for data taken in 2017 (2018). Consequently, these triggers can operate at significantly higher rates compared to the standard triggers. We refer to this approach as ``data scouting''~\cite{Khachatryan:2016ecr}, and the high-rate triggers are termed ``scouting triggers.''

The scouting triggers include an L1 trigger requiring two muons with opposite charge and $\pt > 4\,(4.5)\GeV$ for data collected in 2017 (2018). An additional requirement of $\Delta R < 1.2$ is imposed on the two muons to reduce the trigger rate. Furthermore, an L1 trigger requiring a pair of oppositely charged muons with $\pt > 4.5\GeV$ and $\abs{\eta} < 2.0$, forming an invariant mass between 7 and 18\GeV, is added. These two L1 triggers collect most of the events of interest in the dimuon mass range below $\sim$20\GeV.
Events passing the standard L1 dimuon triggers are also included, and collect a large fraction of events in the dimuon mass range above $\sim$20\GeV. The two muons are required to have $\pt > 3\GeV$ at the HLT. Lower muon thresholds are used for the HLT to maximize efficiency for events accepted by the L1 Trigger. The scouting dimuon triggers were fully commissioned for 2017 and recorded events at a rate of $\sim$2\unit{kHz} at the peak instantaneous luminosity of $2\times10^{34}\percms$. Data corresponding to an integrated luminosity of 96.6\fbinv were collected with these triggers in 2017 and 2018.

The $\PZ_{D}$ signal is simulated at leading order (LO) using the hidden Abelian Higgs model (HAHM v3)~\cite{Gopalakrishna:2008dv,Curtin:2014cca} implemented with the \MGvATNLO 2.2.2 (2.4.2) generator~\cite{Alwall:2014hca} for the 2016 (2017, 2018) data period. The width and cross section of the signal process are directly proportional to $\epsilon^{2}$. The analysis does not rely on simulation for background estimation. However, simulations of certain SM processes are used to evaluate data-to-simulation corrections and corresponding uncertainties in the signal prediction. The process involving the production of the $\Upsilon$(1S) resonance and its decay to a pair of muons is simulated at LO with \PYTHIA 8.230~\cite{Sjostrand:2014zea}. The Drell-Yan (DY) background is simulated at LO and next-to-LO (NLO) using \MGvATNLO, with up to four and two additional partons in the matrix element calculations, respectively.

Events simulated at the matrix element level for the signal and background processes are then interfaced with \PYTHIA in order to simulate the fragmentation, parton shower, and hadronization of partons in the initial and final states, along with the underlying event. This is done using the CUETP8M1 (CP5) tune~\cite{Khachatryan:2015pea,Sirunyan:2019dfx} in the simulation for the 2016 (2017, 2018) data period. Jets from LO (NLO) simulations obtained with \MGvATNLO are matched to the parton shower produced by \PYTHIA following the MLM~\cite{Mangano:2006rw} (FxFx~\cite{Frederix:2012ps}) prescription. The 2016 (2017, 2018) era simulations use the NNPDF 3.0 (3.1) parton distribution functions (PDFs)~\cite{Ball:2014uwa,Ball:2017nwa}. The interactions of all final-state particles with the CMS detector are simulated using \GEANTfour~\cite{Agostinelli:2002hh}. Simulated events include the contribution of particles from additional $\Pp\Pp$ interactions within the same or nearby bunch crossings (pileup), with the multiplicity of reconstructed primary vertices adjusted to match that in data.

Events selected with the standard muon triggers are reconstructed using a particle-flow (PF) algorithm~\cite{Sirunyan:2017ulk} that aims to reconstruct and identify individual particles in an event, with an optimized combination of information from the various elements of the CMS detector. Muons are reconstructed by associating a track reconstructed in the tracking detectors with a track in the muon system. The muon momentum is obtained from the curvature of the corresponding track. The candidate vertex with the largest value of summed physics-object $\pt^2$ is taken to be the primary $\Pp\Pp$ interaction vertex (PV). The physics objects in this case are the jets, clustered using the jet finding algorithm~\cite{Cacciari:2008gp,Cacciari:2011ma} with a distance parameter of 0.4, with the tracks assigned to the candidate vertices as inputs, and the associated missing transverse momentum, taken as the negative vector sum of the \pt of those jets. Jets arising from \PQb quark hadronization and decay (\PQb jets) are identified using a deep neural network algorithm termed ``deepCSV,'' which takes as input tracks that are displaced from the PV, secondary vertices and jet kinematic variables~\cite{Sirunyan:2017ezt}. A working point on the output of the deepCSV algorithm is chosen such that the efficiency of identifying a \PQb jet is $\sim$65--75\% and the probability of misidentifying a light-flavor jet as a \PQb jet is about 1\%.

In the search performed using the standard triggers, events are required to contain at least one well-reconstructed PV and two oppositely charged muons that are geometrically matched to the HLT muon candidates within a cone of $\Delta R=0.1$. The highest and second-highest \pt muons are required to have $\pt > 20$ and 10\GeV, respectively, and $\abs{\eta} < 1.9$. The restriction on $\eta$ is imposed to ensure optimal dimuon mass resolution without incurring a significant loss in acceptance. In events selected with the single-muon triggers, the muon $\pt$ threshold is set to 26\,(29)\GeV for data collected in 2016\,(2017, 2018). The muons are required to pass certain selection requirements based on the quality of their reconstructed tracks. The absolute values of the transverse and longitudinal impact parameters of the muon tracks are required to be less than 0.2 and 0.5\unit{cm}, respectively. The muon isolation, computed as the scalar sum of the \pt of PF candidates (photons, charged hadrons, neutral hadrons) within a cone $\Delta R = 0.4$ around the direction of the muon momentum, is required to be less than 15\% of the muon \pt. Charged PF hadrons not associated with the PV are ignored in this sum. The contribution of neutral particles from pileup, estimated to be one half of the scalar sum of the \pt of charged PF hadrons from pileup in the cone, is subtracted from this sum.
Events containing one or more \PQb jets with $\pt > 30\GeV$ and $\abs{\eta} < 2.4$ are rejected to suppress most of the background from single and pair-produced top quarks.

In the search performed using the scouting triggers, events are required to contain two muons of opposite charge, with $\pt > 4\GeV$ and $\abs{\eta} < 1.9$, that are consistent with originating from the same vertex. The muons are required to pass selection requirements based on the track quality information available in the event. The muon isolation
is required to be less than 15\% of the muon \pt. In order to suppress the background involving muons from heavy flavor decays that typically have low \pt, the (second-) highest \pt muon is required to have $\pt > m_{\PGm\PGm}/(4)~3$, where $m_{\PGm\PGm}$ is the dimuon invariant mass.

\begin{figure}[thb]
    \centering
    \includegraphics[width=.49\textwidth]{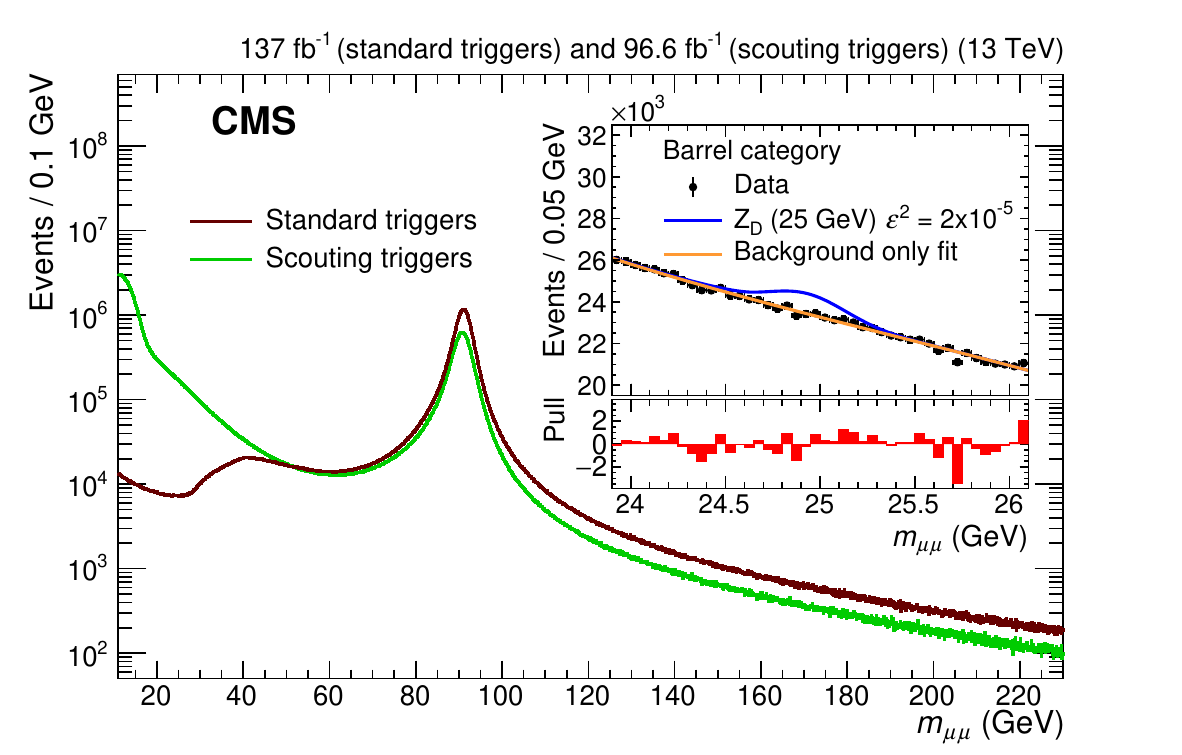}
    \caption{The $m_{\PGm\PGm}$ distributions of events selected with the standard muon triggers (maroon, darker), and the scouting dimuon triggers (green, lighter). Events are required to pass all the selection requirements. The inset is restricted to events in the barrel category in the mass range 23.9--26.1\GeV. A function describing the background is fit to these data, and a 25\GeV $\PZ_{D}$ signal corresponding to $\epsilon^{2} = 2 \times 10^{-5}$ is added. The bottom panel of the inset shows the bin-by-bin difference between the number of events in data and the prediction from the background fit, divided by the statistical uncertainty.}
    \label{fig:ScoutingDimuonMass}
\end{figure}

Figure~\ref{fig:ScoutingDimuonMass} shows the $m_{\PGm\PGm}$ distributions obtained with data collected using the standard and scouting triggers. The $m_{\PGm\PGm}$ distribution of fully reconstructed events collected using the standard triggers suffers from a significant acceptance loss resulting in the structure visible for $m_{\PGm\PGm} \lesssim40\GeV$. The loss of acceptance is due to the \pt thresholds of 20 and 10\GeV on the two muons. The $m_{\PGm\PGm}$ distribution of events collected using the scouting triggers, however, continues to rise steadily for masses lower than 40\GeV. For masses lower than 20\GeV, an even steeper rise is observed. This is the region where the L1 trigger with a requirement for $m_{\PGm\PGm}$ to be in the range 7--18\GeV contributes significantly.

The $m_{\PGm\PGm}$ resolution depends strongly on the $\abs{\eta}$ of the two muons. The \pt resolution of muons with $\pt < 50\GeV$ is $\sim$1\% in the central barrel region of the detector ($\abs{\eta} < 0.9$), and $\sim$3\% in the end caps of the muon system ($\abs{\eta} > 1.2$)~\cite{Sirunyan:2018fpa}.
Therefore, events are divided in two categories. The barrel category consists of events in which both muons have $\abs{\eta} < 0.9$, and the forward category contains events in which at least one of the two muons has $0.9 < \abs{\eta} < 1.9$.

The intrinsic width of the signal resonance considered in this search is assumed to be much narrower than the detector resolution. The signal line shape is modeled using a parametric shape called the double-sided crystal ball (DCB) function~\cite{Gaiser:1982}. The core of this function consists of a Gaussian distribution of mean $s$ and standard deviation $\sigma$. The $s$ and $\sigma$ parameters represent the peak position and resolution of the resonance, and are allowed to vary using constrained nuisance parameters corresponding to the muon momentum scale and resolution uncertainties. The uncertainty in the $m_{\PGm\PGm}$ resolution is estimated to be 10\% of $\sigma$. The uncertainty in the mass scale is 0.1\% of the resonance mass in the search performed using data from standard triggers, while it varies in the range 0.06--0.1\% for the search performed using data from the scouting triggers.

There are several experimental sources of systematic uncertainty in the estimation of the signal yield. The measured integrated luminosity has an uncertainty of 2.3 (2.5)\% for data collected in 2017 (2016, 2018)~\cite{CMS-PAS-LUM-17-001,CMS-PAS-LUM-17-004,CMS-PAS-LUM-18-002}. In the search performed using the standard muon triggers, uncertainties of 0.2\% and 1.5--3.0\% are ascribed to the measurement of the trigger and muon selection efficiencies, respectively. The uncertainty in the inefficiency caused by the \PQb jet rejection is 1\%. In the search performed using scouting data, the uncertainty in the muon selection efficiency varies in the range 3--10\%. The efficiency of the scouting triggers is measured with respect to fully reconstructed muons using events selected with independent standard triggers. The average uncertainty in the trigger efficiency is of order 5\%, a value that takes into account a small uncertainty associated with the requirement that the muons be fully reconstructed, and a potential effect arising from background contamination.

In order to extract the signal from data, a simultaneous binned maximum likelihood fit is performed to the $m_{\PGm\PGm}$ distributions in the barrel and forward event categories. A parametric function is used to model the shape of the background. The parameters of this function and the background yield are allowed to float freely in the fit, and are uncorrelated between the two event categories. In the search using the standard triggers, the fit is performed in a range of $\pm 7 \sigma$ around the probed resonance mass, where $\sigma$ is the mass resolution parameter of the DCB function used to model the signal. The background, which is dominated by DY events, is modeled in both event categories using the product of a modified form of the Breit-Wigner function~\cite{Sirunyan:2018hbu} and a second-order Bernstein polynomial. In the search using the scouting triggers, the fit is performed in a range of $\pm 5 \sigma$ around the probed resonance mass. The background mass distribution is modeled using a fourth-order Bernstein polynomial. The chosen background functions, and the sizes of the mass windows maximize the expected sensitivity while introducing only a small bias in the measured signal yield. Figure~\ref{fig:ScoutingDimuonMass}~(inset) shows, as an example, the $m_{\PGm\PGm}$ distribution of events in the barrel category in the 23.9--26.1\GeV window that is used to search for a 25\GeV resonance. A fit performed to these data assuming no signal is also shown along with the expected signal contribution for a 25\GeV $\PZ_{D}$ with $\epsilon^{2} = 2 \times 10^{-5}$.

In order to assess the possible bias in the measurement of a signal due to the choice of the function used to model the background distribution, we consider several alternative functions to model the shape of the background. An alternative test function is fit to the data in each event category, and is then used to generate 2000 sets of pseudodata. A certain amount of signal is injected in each of the pseudodata sets. A signal-plus-background fit is performed using the chosen background function to each of the pseudodata sets with the signal yield allowed to float freely. The bias is quantified as the ratio of the difference between the measured and injected signals, and the statistical uncertainty in the measured signal yield. The bias is found to be less than 10\% of the statistical uncertainty in the measured signal yield for resonance masses greater than 20\GeV, in both event categories. For masses below 20\GeV, the bias is estimated to be less than 20\,(30)\% of the statistical uncertainty for the barrel (forward) category. In order to account for this uncertainty, we introduce an additional background in the fit. The shape of the background is taken to be the same as that of the signal, and its normalization is treated as a nuisance parameter with mean value zero, whose variations are constrained by a Gaussian distribution with standard deviation equal to the bias found in the tests.

The statistical analysis is performed using a profile likelihood ratio test statistic in which systematic uncertainties are modeled as constrained nuisance parameters. The nuisance parameters for the $m_{\PGm\PGm}$ scale and resolution uncertainties, and the bias due to the choice of the background fit function are modeled using the Gaussian probability density function. All other sources of systematic uncertainty affect the signal yield, and are modeled with nuisance parameters that are constrained using the log-normal probability density function.

\begin{figure}[thbp]
    \centering
    \includegraphics[width=.49\textwidth]{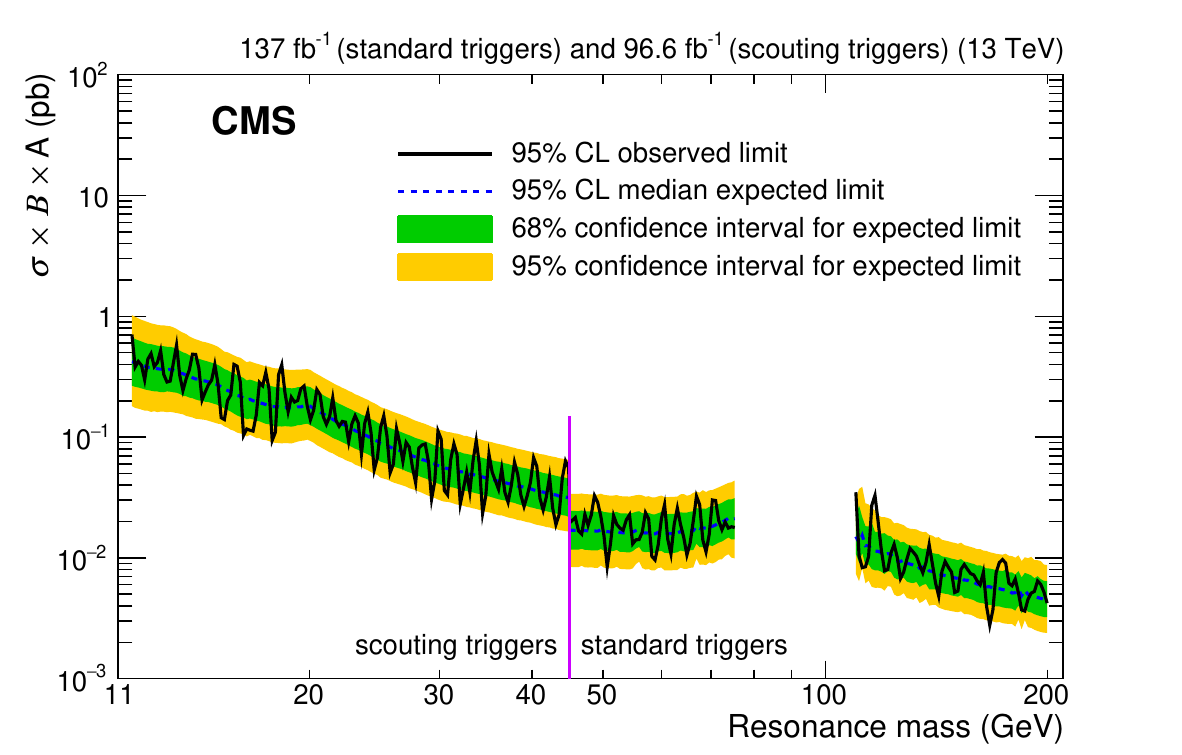}
    \caption{Expected and observed upper limits at 95\% \CL on the product of the signal cross section ($\sigma$) for a narrow resonance, branching fraction to a pair of muons ($\mathcal{B}$), and acceptance ($A$) as a function of the mass of a narrow resonance. Results obtained using the scouting (standard) triggers are to the left (right) of the vertical purple line.}
    \label{fig:Limits}
\end{figure}

The data are found to be consistent with the background expectation. Figure~\ref{fig:Limits} shows the upper limits at 95\% confidence level (\CL) on the product of the signal cross section, branching fraction to a pair of muons, and the kinematic and geometrical acceptance for a narrow resonance using an asymptotic \CLs criterion~\cite{CLS1,CLS2,Cowan:2010js}.

\begin{figure}[thb]
    \centering
    \includegraphics[width=.49\textwidth]{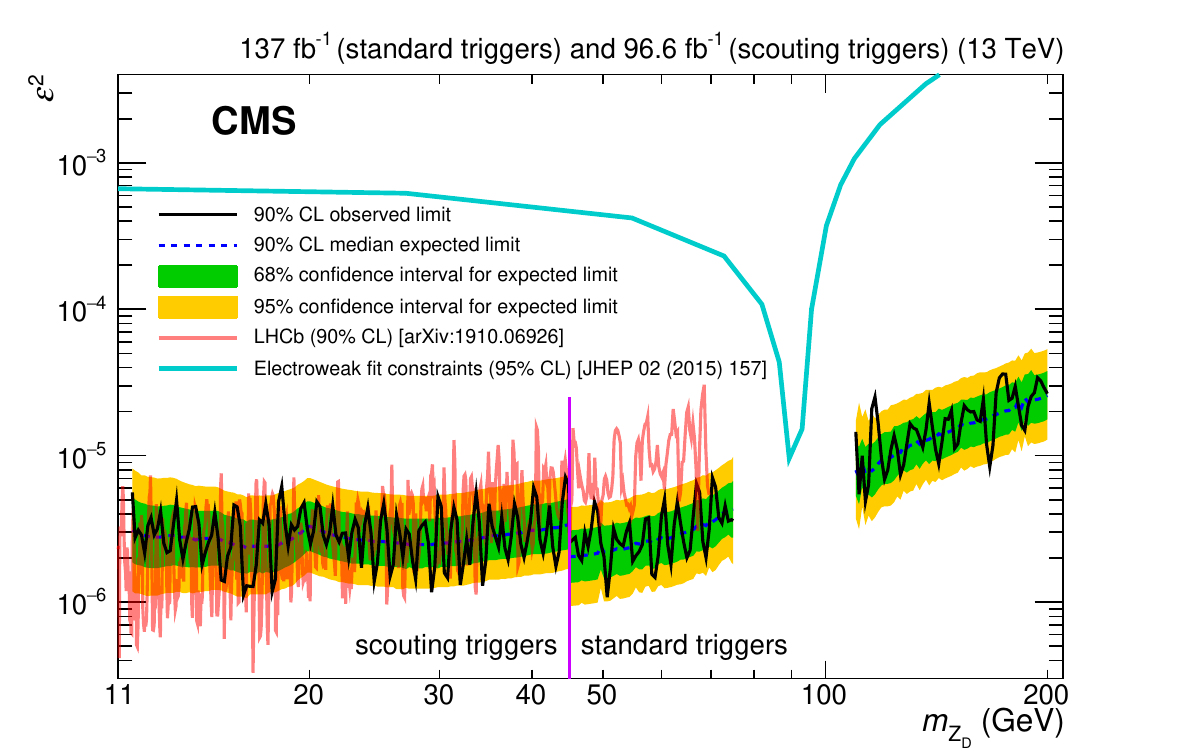}
    \caption{Expected and observed upper limits at 90\% \CL on $\epsilon^{2}$, the square of the kinetic mixing coefficient, as a function of the $\PZ_{D}$ mass. Results obtained using the scouting (standard) triggers are to the left (right) of the vertical purple line. Limits at 90\% \CL from the search performed by the LHCb Collaboration~\cite{Aaij:2019bvg} are shown in red, and constraints at 95\% \CL from the measurements of the electroweak observables are shown in light blue~\cite{Curtin:2014cca}.}
    \label{fig:EpsilonLimits}
\end{figure}

The results of this search are interpreted in the context of a dark photon model. The signal is kinematically similar to the SM DY process. Therefore, an NLO $K$ factor has been computed for the LO signal simulation by comparing the NLO and LO DY cross sections, and is found to be 1.09 over the entire mass range under consideration. Uncertainties due to the choice of the renormalization and factorization scales (4.5\%), and the modeling of the PDFs (1\%) are ascribed to the signal cross section. We set upper limits at 90\% \CL on $\epsilon^{2}$ as a function of the $\PZ_{D}$ mass, as shown in Fig.~\ref{fig:EpsilonLimits}. These are compared with recent results from the LHCb Collaboration~\cite{Aaij:2019bvg,Ilten:2018crw} and indirect constraints at 95\% \CL from measurements of the electroweak observables~\cite{Curtin:2014cca}. This search sets the most stringent limits to date in the $\sim$30--75 and 110--200\GeV mass ranges. Furthermore, limits from this search are competitive with those obtained in Ref.~\cite{Aaij:2019bvg} at lower masses.

In summary, a search has been presented for a narrow resonance decaying to a pair of muons using proton-proton collision data recorded by the CMS experiment at $\sqrt{s}=13\TeV$. The search in the 45--75 and 110--200\GeV resonance mass ranges uses fully reconstructed data containing a pair of muons with transverse momenta greater than 20 and 10\GeV, corresponding to an integrated luminosity of 137\fbinv.  The search in the resonance mass range of 11.5--45.0\GeV is performed using data collected with high-rate dimuon triggers, corresponding to an integrated luminosity of 96.6\fbinv. This is the first search that uses data with reduced trigger-level muon information, collected with dimuon triggers that have transverse momentum thresholds of 3\GeV. The data are found to be consistent with the background prediction. The search sets the lowest upper limits to date on the kinetic mixing coefficient of a dark photon in the $\sim$30--75 and 110--200\GeV mass ranges.

\begin{acknowledgments}
We congratulate our colleagues in the CERN accelerator departments for the excellent performance of the LHC and thank the technical and administrative staffs at CERN and at other CMS institutes for their contributions to the success of the CMS effort. In addition, we gratefully acknowledge the computing centers and personnel of the Worldwide LHC Computing Grid for delivering so effectively the computing infrastructure essential to our analyses. Finally, we acknowledge the enduring support for the construction and operation of the LHC and the CMS detector provided by the following funding agencies: BMBWF and FWF (Austria); FNRS and FWO (Belgium); CNPq, CAPES, FAPERJ, FAPERGS, and FAPESP (Brazil); MES (Bulgaria); CERN; CAS, MoST, and NSFC (China); COLCIENCIAS (Colombia); MSES and CSF (Croatia); RPF (Cyprus); SENESCYT (Ecuador); MoER, ERC IUT, PUT and ERDF (Estonia); Academy of Finland, MEC, and HIP (Finland); CEA and CNRS/IN2P3 (France); BMBF, DFG, and HGF (Germany); GSRT (Greece); NKFIA (Hungary); DAE and DST (India); IPM (Iran); SFI (Ireland); INFN (Italy); MSIP and NRF (Republic of Korea); MES (Latvia); LAS (Lithuania); MOE and UM (Malaysia); BUAP, CINVESTAV, CONACYT, LNS, SEP, and UASLP-FAI (Mexico); MOS (Montenegro); MBIE (New Zealand); PAEC (Pakistan); MSHE and NSC (Poland); FCT (Portugal); JINR (Dubna); MON, RosAtom, RAS, RFBR, and NRC KI (Russia); MESTD (Serbia); SEIDI, CPAN, PCTI, and FEDER (Spain); MOSTR (Sri Lanka); Swiss Funding Agencies (Switzerland); MST (Taipei); ThEPCenter, IPST, STAR, and NSTDA (Thailand); TUBITAK and TAEK (Turkey); NASU (Ukraine); STFC (United Kingdom); DOE and NSF (USA). 
\end{acknowledgments}

\bibliography{auto_generated}

\providecommand{\href}[2]{#2}\begingroup\raggedright\begin{thebibliography}{10}%
\makeatletter
\providecommand{\hrefCMSnoop }[0]{\@secondoftwo}%
\makeatother
\providecommand{\doi}{\texttt{doi:}\begingroup \urlstyle{tt}\Url}

\bibitem{Bertone:2010zza}
\hrefCMSnoop {}{G.~Bertone and J.~Silk, ``{Particle Dark Matter: Observations,
  Models and Searches}''}.
\newblock Cambridge University Press, Cambridge, England, 2010.

\bibitem{Feng:2010gw}
\hrefCMSnoop {}{J.~L. Feng, ``{Dark matter candidates from particle physics and
  methods of detection}'',} \textit{ Annu. Rev. Astron. Astrophys.} \textbf{
  48} (2010) 495,
  \href{http://dx.doi.org/10.1146/annurev-astro-082708-101659}{\doi{10.1146/annurev-astro-082708-101659}},
\href{http://www.arXiv.org/abs/1003.0904}{\texttt{arXiv:1003.0904}}.

\bibitem{Porter:2011nv}
\hrefCMSnoop {}{T.~A. Porter, R.~P. Johnson, and P.~W. Graham, ``{Dark matter
  searches with astroparticle data}'',} \textit{ Annu. Rev. Astron. Astrophys.}
  \textbf{ 49} (2011) 155,
  \href{http://dx.doi.org/10.1146/annurev-astro-081710-102528}{\doi{10.1146/annurev-astro-081710-102528}},
\href{http://www.arXiv.org/abs/1104.2836}{\texttt{arXiv:1104.2836}}.

\bibitem{Ade:2015xua}
\hrefCMSnoop {}{{Planck} Collaboration, ``{Planck 2015 results. XIII.
  Cosmological parameters}'',} \textit{ Astron. Astrophys.} \textbf{ 594}
  (2016) A13,
  \href{http://dx.doi.org/10.1051/0004-6361/201525830}{\doi{10.1051/0004-6361/201525830}},
\href{http://www.arXiv.org/abs/1502.01589}{\texttt{arXiv:1502.01589}}.

\bibitem{Essig:2013lka}
\hrefCMSnoop {}{R.~Essig {et~al.}, ``Dark sectors and new, light weakly-coupled
  particles'',} in \textit{ Planning the Future of {U.S.} Particle Physics: The
  {Snowmass} 2013 Proceedings}, R.~H. Bernstein {et~al.}, eds.
\newblock Minneapolis, MN, USA,
July-August, 2013.
\newblock

\bibitem{Galison:1983pa}
\hrefCMSnoop {}{P.~Galison and A.~Manohar, ``{Two Z's or not two Z's?}'',}
  \textit{ Phys. Lett. B} \textbf{ 136} (1984) 279,
\href{http://dx.doi.org/10.1016/0370-2693(84)91161-4}{\doi{10.1016/0370-2693(84)91161-4}}.

\bibitem{Holdom:1985ag}
\hrefCMSnoop {}{B.~Holdom, ``{Two U(1)'s and $\epsilon$ charge shifts}'',}
  \textit{ Phys. Lett. B} \textbf{ 166} (1986) 196,
\href{http://dx.doi.org/10.1016/0370-2693(86)91377-8}{\doi{10.1016/0370-2693(86)91377-8}}.

\bibitem{Dienes:1996zr}
\hrefCMSnoop {}{K.~R. Dienes, C.~F. Kolda, and J.~March-Russell, ``{Kinetic
  mixing and the supersymmetric gauge hierarchy}'',} \textit{ Nucl. Phys. B}
  \textbf{ 492} (1997) 104,
  \href{http://dx.doi.org/10.1016/S0550-3213(97)00173-9}{\doi{10.1016/S0550-3213(97)00173-9}},
\href{http://www.arXiv.org/abs/hep-ph/9610479}{\texttt{arXiv:hep-ph/9610479}}.

\bibitem{Curtin:2014cca}
\hrefCMSnoop {}{D.~Curtin, R.~Essig, S.~Gori, and J.~Shelton, ``{Illuminating
  dark photons with high-energy colliders}'',} \textit{ JHEP} \textbf{ 02}
  (2015) 157,
  \href{http://dx.doi.org/10.1007/JHEP02(2015)157}{\doi{10.1007/JHEP02(2015)157}},
\href{http://www.arXiv.org/abs/1412.0018}{\texttt{arXiv:1412.0018}}.

\bibitem{Konaka:1986cb}
\hrefCMSnoop {}{A.~Konaka {et~al.}, ``{Search for Neutral Particles in
  Electron-Beam-Dump Experiment}'',} \textit{ Phys. Rev. Lett.} \textbf{ 57}
  (1986) 659,
\href{http://dx.doi.org/10.1103/PhysRevLett.57.659}{\doi{10.1103/PhysRevLett.57.659}}.

\bibitem{Abrahamyan:2011gv}
\hrefCMSnoop {}{{APEX} Collaboration, ``{Search for a New Gauge Boson in
  Electron-Nucleus Fixed-Target Scattering by the APEX Experiment}'',} \textit{
  Phys. Rev. Lett.} \textbf{ 107} (2011) 191804,
  \href{http://dx.doi.org/10.1103/PhysRevLett.107.191804}{\doi{10.1103/PhysRevLett.107.191804}},
\href{http://www.arXiv.org/abs/1108.2750}{\texttt{arXiv:1108.2750}}.

\bibitem{MeijerDrees:1992kd}
\hrefCMSnoop {}{{SINDRUM I} Collaboration, ``{Search for Weakly Interacting
  Neutral Bosons Produced in $\pi^{-}$p Interactions at Rest and Decaying Into
  e$^{+}$e$^{-}$ Pairs}'',} \textit{ Phys. Rev. Lett.} \textbf{ 68} (1992)
  3845,
\href{http://dx.doi.org/10.1103/PhysRevLett.68.3845}{\doi{10.1103/PhysRevLett.68.3845}}.

\bibitem{Lees:2014xha}
\hrefCMSnoop {}{{BaBar} Collaboration, ``{Search for a Dark Photon in $e^+e^-$
  Collisions at BaBar}'',} \textit{ Phys. Rev. Lett.} \textbf{ 113} (2014)
  201801,
  \href{http://dx.doi.org/10.1103/PhysRevLett.113.201801}{\doi{10.1103/PhysRevLett.113.201801}},
\href{http://www.arXiv.org/abs/1406.2980}{\texttt{arXiv:1406.2980}}.

\bibitem{Ilten:2016tkc}
P.~Ilten\hrefCMSnoop {}{ {et~al.}, ``{Proposed Inclusive Dark Photon Search at
  LHCb}'',} \textit{ Phys. Rev. Lett.} \textbf{ 116} (2016) 251803,
  \href{http://dx.doi.org/10.1103/PhysRevLett.116.251803}{\doi{10.1103/PhysRevLett.116.251803}},
\href{http://www.arXiv.org/abs/1603.08926}{\texttt{arXiv:1603.08926}}.

\bibitem{Aaij:2017rft}
\hrefCMSnoop {}{{LHCb Collaboration}, ``{Search for Dark Photons Produced in 13
  TeV $pp$ Collisions}'',} \textit{ Phys. Rev. Lett.} \textbf{ 120} (2018)
  061801,
  \href{http://dx.doi.org/10.1103/PhysRevLett.120.061801}{\doi{10.1103/PhysRevLett.120.061801}},
\href{http://www.arXiv.org/abs/1710.02867}{\texttt{arXiv:1710.02867}}.

\bibitem{Aaij:2019bvg}
\hrefCMSnoop {}{{LHCb Collaboration}, ``{Search for $A'\!\to\!\mu^+\mu^-$
  Decays}'',} \textit{ Phys. Rev. Lett.} \textbf{ 124} (2020) 041801,
  \href{http://dx.doi.org/10.1103/PhysRevLett.124.041801}{\doi{10.1103/PhysRevLett.124.041801}},
\href{http://www.arXiv.org/abs/1910.06926}{\texttt{arXiv:1910.06926}}.

\bibitem{Chatrchyan:2008zzk}
\hrefCMSnoop {}{{CMS Collaboration}, ``The {CMS} experiment at the {CERN}
  {LHC}'',} \textit{ JINST} \textbf{ 3} (2008) S08004,
  \href{http://dx.doi.org/10.1088/1748-0221/3/08/S08004}{\doi{10.1088/1748-0221/3/08/S08004}}.

\bibitem{Khachatryan:2016bia}
\hrefCMSnoop {}{{CMS Collaboration}, ``{The CMS trigger system}'',} \textit{
  JINST} \textbf{ 12} (2017) P01020,
  \href{http://dx.doi.org/10.1088/1748-0221/12/01/P01020}{\doi{10.1088/1748-0221/12/01/P01020}},
\href{http://www.arXiv.org/abs/1609.02366}{\texttt{arXiv:1609.02366}}.

\bibitem{Khachatryan:2016ecr}
\hrefCMSnoop {}{{CMS Collaboration}, ``{Search for Narrow Resonances in Dijet
  Final States at $\sqrt{s}=$ 8 TeV With the Novel CMS Technique of Data
  Scouting}'',} \textit{ Phys. Rev. Lett.} \textbf{ 117} (2016) 031802,
  \href{http://dx.doi.org/10.1103/PhysRevLett.117.031802}{\doi{10.1103/PhysRevLett.117.031802}},
\href{http://www.arXiv.org/abs/1604.08907}{\texttt{arXiv:1604.08907}}.

\bibitem{Gopalakrishna:2008dv}
\hrefCMSnoop {}{S.~Gopalakrishna, S.~Jung, and J.~D. Wells, ``{Higgs boson
  decays to four fermions through an Abelian hidden sector}'',} \textit{ Phys.
  Rev. D} \textbf{ 78} (2008) 055002,
  \href{http://dx.doi.org/10.1103/PhysRevD.78.055002}{\doi{10.1103/PhysRevD.78.055002}},
\href{http://www.arXiv.org/abs/0801.3456}{\texttt{arXiv:0801.3456}}.

\bibitem{Alwall:2014hca}
J.~Alwall\hrefCMSnoop {}{ {et~al.}, ``{The automated computation of tree-level
  and next-to-leading order differential cross sections, and their matching to
  parton shower simulations}'',} \textit{ JHEP} \textbf{ 07} (2014) 079,
  \href{http://dx.doi.org/10.1007/JHEP07(2014)079}{\doi{10.1007/JHEP07(2014)079}},
\href{http://www.arXiv.org/abs/1405.0301}{\texttt{arXiv:1405.0301}}.

\bibitem{Sjostrand:2014zea}
T.~Sj{\"o}strand\hrefCMSnoop {}{ {et~al.}, ``An introduction to {PYTHIA
  8.2}'',} \textit{ Comput. Phys. Commun.} \textbf{ 191} (2015) 159,
  \href{http://dx.doi.org/10.1016/j.cpc.2015.01.024}{\doi{10.1016/j.cpc.2015.01.024}},
\href{http://www.arXiv.org/abs/1410.3012}{\texttt{arXiv:1410.3012}}.

\bibitem{Khachatryan:2015pea}
\hrefCMSnoop {}{{CMS Collaboration}, ``{Event generator tunes obtained from
  underlying event and multiparton scattering measurements}'',} \textit{ Eur.
  Phys. J. C} \textbf{ 76} (2016) 155,
  \href{http://dx.doi.org/10.1140/epjc/s10052-016-3988-x}{\doi{10.1140/epjc/s10052-016-3988-x}},
\href{http://www.arXiv.org/abs/1512.00815}{\texttt{arXiv:1512.00815}}.

\bibitem{Sirunyan:2019dfx}
\hrefCMSnoop {}{{CMS Collaboration}, ``{Extraction and validation of a new set
  of CMS PYTHIA8 tunes from underlying-event measurements}'',} \textit{ Eur.
  Phys. J. C} \textbf{ 80} (2020) 4,
  \href{http://dx.doi.org/10.1140/epjc/s10052-019-7499-4}{\doi{10.1140/epjc/s10052-019-7499-4}},
\href{http://www.arXiv.org/abs/1903.12179}{\texttt{arXiv:1903.12179}}.

\bibitem{Mangano:2006rw}
\hrefCMSnoop {}{M.~L. Mangano, M.~Moretti, F.~Piccinini, and M.~Treccani,
  ``{Matching matrix elements and shower evolution for top-pair production in
  hadronic collisions}'',} \textit{ JHEP} \textbf{ 01} (2007) 013,
  \href{http://dx.doi.org/10.1088/1126-6708/2007/01/013}{\doi{10.1088/1126-6708/2007/01/013}},
\href{http://www.arXiv.org/abs/hep-ph/0611129}{\texttt{arXiv:hep-ph/0611129}}.

\bibitem{Frederix:2012ps}
\hrefCMSnoop {}{R.~Frederix and S.~Frixione, ``{Merging meets matching in
  MC@NLO}'',} \textit{ JHEP} \textbf{ 12} (2012) 061,
  \href{http://dx.doi.org/10.1007/JHEP12(2012)061}{\doi{10.1007/JHEP12(2012)061}},
\href{http://www.arXiv.org/abs/1209.6215}{\texttt{arXiv:1209.6215}}.

\bibitem{Ball:2014uwa}
\hrefCMSnoop {}{{NNPDF} Collaboration, ``{Parton distributions for the LHC Run
  II}'',} \textit{ JHEP} \textbf{ 04} (2015) 040,
  \href{http://dx.doi.org/10.1007/JHEP04(2015)040}{\doi{10.1007/JHEP04(2015)040}},
\href{http://www.arXiv.org/abs/1410.8849}{\texttt{arXiv:1410.8849}}.

\bibitem{Ball:2017nwa}
\hrefCMSnoop {}{{NNPDF} Collaboration, ``{Parton distributions from
  high-precision collider data}'',} \textit{ Eur. Phys. J. C} \textbf{ 77}
  (2017) 663,
  \href{http://dx.doi.org/10.1140/epjc/s10052-017-5199-5}{\doi{10.1140/epjc/s10052-017-5199-5}},
\href{http://www.arXiv.org/abs/1706.00428}{\texttt{arXiv:1706.00428}}.

\bibitem{Agostinelli:2002hh}
\hrefCMSnoop {}{{GEANT4} Collaboration, ``{{\GEANTfour}---A simulation
  toolkit}'',} \textit{ Nucl. Instrum. Methods Phys. Res., Sect. A} \textbf{
  506} (2003) 250,
\href{http://dx.doi.org/10.1016/S0168-9002(03)01368-8}{\doi{10.1016/S0168-9002(03)01368-8}}.

\bibitem{Sirunyan:2017ulk}
\hrefCMSnoop {}{{CMS Collaboration}, ``{Particle-flow reconstruction and global
  event description with the CMS detector}'',} \textit{ JINST} \textbf{ 12}
  (2017) P10003,
  \href{http://dx.doi.org/10.1088/1748-0221/12/10/P10003}{\doi{10.1088/1748-0221/12/10/P10003}},
\href{http://www.arXiv.org/abs/1706.04965}{\texttt{arXiv:1706.04965}}.

\bibitem{Cacciari:2008gp}
\hrefCMSnoop {}{M.~Cacciari, G.~P. Salam, and G.~Soyez, ``The anti-\kt jet
  clustering algorithm'',} \textit{ JHEP} \textbf{ 04} (2008) 063,
  \href{http://dx.doi.org/10.1088/1126-6708/2008/04/063}{\doi{10.1088/1126-6708/2008/04/063}},
  \href{http://www.arXiv.org/abs/0802.1189}{\texttt{arXiv:0802.1189}}.

\bibitem{Cacciari:2011ma}
\hrefCMSnoop {}{M.~Cacciari, G.~P. Salam, and G.~Soyez, ``{FastJet user
  manual}'',} \textit{ Eur. Phys. J. C} \textbf{ 72} (2012) 1896,
  \href{http://dx.doi.org/10.1140/epjc/s10052-012-1896-2}{\doi{10.1140/epjc/s10052-012-1896-2}},
\href{http://www.arXiv.org/abs/1111.6097}{\texttt{arXiv:1111.6097}}.

\bibitem{Sirunyan:2017ezt}
\hrefCMSnoop {}{{CMS Collaboration}, ``{Identification of heavy-flavour jets
  with the CMS detector in pp collisions at 13 TeV}'',} \textit{ JINST}
  \textbf{ 13} (2018) P05011,
  \href{http://dx.doi.org/10.1088/1748-0221/13/05/P05011}{\doi{10.1088/1748-0221/13/05/P05011}},
\href{http://www.arXiv.org/abs/1712.07158}{\texttt{arXiv:1712.07158}}.

\bibitem{Sirunyan:2018fpa}
\hrefCMSnoop {}{{CMS Collaboration}, ``{Performance of the CMS muon detector
  and muon reconstruction with proton-proton collisions at $\sqrt{s}=$ 13
  TeV}'',} \textit{ JINST} \textbf{ 13} (2018) P06015,
  \href{http://dx.doi.org/10.1088/1748-0221/13/06/P06015}{\doi{10.1088/1748-0221/13/06/P06015}},
\href{http://www.arXiv.org/abs/1804.04528}{\texttt{arXiv:1804.04528}}.

\bibitem{Gaiser:1982}
\href
  {https://www-public.slac.stanford.edu/sciDoc/docMeta.aspx?slacPubNumber=slac-r-255}{J.~E.
  Gaiser, ``Charmonium spectroscopy from radiative decays of the {$J/\psi$ and
  $\psi'$}''}.
\newblock PhD thesis, Stanford University, 1982.
\newblock {SLAC} Report {SLAC-R-255}.

\bibitem{CMS-PAS-LUM-17-001}
\href {https://cds.cern.ch/record/2257069}{{CMS Collaboration}, ``{CMS}
  luminosity measurements for the 2016 data taking period'',} Technical Report
  CMS-PAS-LUM-17-001, CERN, Geneva, 2017.

\bibitem{CMS-PAS-LUM-17-004}
\href {https://cds.cern.ch/record/2621960}{{CMS Collaboration}, ``{CMS}
  luminosity measurement for the 2017 data-taking period at $\sqrt{s} =
  13~\mathrm{TeV}$'',} Technical Report CMS-PAS-LUM-17-004, CERN, Geneva, 2018.

\bibitem{CMS-PAS-LUM-18-002}
\href {https://cds.cern.ch/record/2676164}{{CMS Collaboration}, ``{CMS
  luminosity measurement for the 2018 data-taking period at $\sqrt{s} =
  13~\mathrm{TeV}$}'',} Technical Report CMS-PAS-LUM-18-002, CERN, Geneva,
  2019.

\bibitem{Sirunyan:2018hbu}
\hrefCMSnoop {}{{CMS Collaboration}, ``{Search for the Higgs Boson Decaying to
  Two Muons in Proton-Proton Collisions at $\sqrt{s} =$ 13 TeV}'',} \textit{
  Phys. Rev. Lett.} \textbf{ 122} (2019) 021801,
  \href{http://dx.doi.org/10.1103/PhysRevLett.122.021801}{\doi{10.1103/PhysRevLett.122.021801}},
\href{http://www.arXiv.org/abs/1807.06325}{\texttt{arXiv:1807.06325}}.

\bibitem{CLS1}
\hrefCMSnoop {}{A.~L. Read, ``Presentation of search results: The
  {CL$_{\text{s}}$} technique'',} \textit{ J. Phys. G} \textbf{ 28} (2002)
  2693,
\href{http://dx.doi.org/10.1088/0954-3899/28/10/313}{\doi{10.1088/0954-3899/28/10/313}}.

\bibitem{CLS2}
\hrefCMSnoop {}{T.~Junk, ``Confidence level computation for combining searches
  with small statistics'',} \textit{ Nucl. Instrum. Methods Phys. Res., Sect.
  A} \textbf{ 434} (1999) 435,
  \href{http://dx.doi.org/10.1016/S0168-9002(99)00498-2}{\doi{10.1016/S0168-9002(99)00498-2}},
\href{http://www.arXiv.org/abs/hep-ex/9902006}{\texttt{arXiv:hep-ex/9902006}}.

\bibitem{Cowan:2010js}
\hrefCMSnoop {}{G.~Cowan, K.~Cranmer, E.~Gross, and O.~Vitells, ``Asymptotic
  formulae for likelihood-based tests of new physics'',} \textit{ Eur. Phys. J.
  C} \textbf{ 71} (2011) 1554,
  \href{http://dx.doi.org/10.1140/epjc/s10052-011-1554-0}{\doi{10.1140/epjc/s10052-011-1554-0}},
\href{http://www.arXiv.org/abs/1007.1727}{\texttt{arXiv:1007.1727}}.

\bibitem{Ilten:2018crw}
\hrefCMSnoop {}{P.~Ilten, Y.~Soreq, M.~Williams, and W.~Xue, ``{Serendipity in
  dark photon searches}'',} \textit{ JHEP} \textbf{ 06} (2018) 004,
  \href{http://dx.doi.org/10.1007/JHEP06(2018)004}{\doi{10.1007/JHEP06(2018)004}},
\href{http://www.arXiv.org/abs/1801.04847}{\texttt{arXiv:1801.04847}}.

\end{thebibliography}\endgroup
\cleardoublepage \appendix\section{The CMS Collaboration \label{app:collab}}\begin{sloppypar}\hyphenpenalty=5000\widowpenalty=500\clubpenalty=5000\vskip\cmsinstskip
\textbf{Yerevan Physics Institute, Yerevan, Armenia}\\*[0pt]
A.M.~Sirunyan$^{\textrm{\dag}}$, A.~Tumasyan
\vskip\cmsinstskip
\textbf{Institut f\"{u}r Hochenergiephysik, Wien, Austria}\\*[0pt]
W.~Adam, F.~Ambrogi, T.~Bergauer, M.~Dragicevic, J.~Er\"{o}, A.~Escalante~Del~Valle, M.~Flechl, R.~Fr\"{u}hwirth\cmsAuthorMark{1}, M.~Jeitler\cmsAuthorMark{1}, N.~Krammer, I.~Kr\"{a}tschmer, D.~Liko, T.~Madlener, I.~Mikulec, N.~Rad, J.~Schieck\cmsAuthorMark{1}, R.~Sch\"{o}fbeck, M.~Spanring, D.~Spitzbart, W.~Waltenberger, C.-E.~Wulz\cmsAuthorMark{1}, M.~Zarucki
% ...
\vskip\cmsinstskip
\dag: Deceased\\
1:  Also at Vienna University of Technology, Vienna, Austria\\
% ...
\end{sloppypar}
\end{document}